\documentstyle[preprint,prd,aps]{revtex}
\begin{document}
\draft

\title{$\bf Sp(4,R)/GL(2,R)$ Matrix Structure of Geodesic Solutions
for Einstein--Maxwell--Dilaton--Axion Theory}

\author{Kechkin Oleg and Yurova Maria }

\address{Nuclear Physics Institute,\\
Moscow State University, \\
Moscow 119899, RUSSIA, \\
e-mail: kechkin@cdfe.npi.msu.su}

\date{\today}

\maketitle

\draft

\begin{abstract}
The constructed $Sp(4,R)/GL(2,R)$ matrix operator defines the family of
isotropic geodesic containing vacuum point lines in the
target space of the stationary D=4 Einstein--Maxwell--dilaton--axion theory.
This operator is used to derive a class of solutions which describes a
point center system with nontrivial values of mass, parameter NUT, as well as
electric, magnetic, dilaton and axion charges. It is shown that this class
contains both particular solutions Majumdar--Papapetrou--like black holes
and massless asymptotically nonflat naked singularities.
\end{abstract}


\draft

\narrowtext

\section{Introduction}
In recent years much attention has been given to the study of gravitational
models appearing in superstring theory low energy limit
\cite {gsw}--\cite {mah}.
Einstein--Maxwell
theory with dilaton and axion fields (EMDA) is one of such models.
It appears in the
frames of heterotic string theory as a result of omission of
a part of vector and
scalar fields arising during extra dimensions compactification. As it has
been established earlier, the theory under consideration leads to
three--dimensional $\sigma$--model with symmetric target space which
possesses an isometry group locally isomorphic to $Sp(4,R)$
\cite {jmp}, \cite {pr1}, and the model
admits a null--curvature $Sp(4,R)/U(2)$ coset representation
\cite {prl}--\cite {pl}. The brief
description of matrix formalism is given in the following section.

Subsequently the class of exact solutions to the motion equations
written in matrix
form is constructed. Using the Kramer--Neugebauer approach \cite {kn},
we consider coset space matrix $M$
dependance on one space coordinate function $\lambda (x^i)$.
The found solutions are corresponding to
isotropic geodesic lines family in the target space and to the set of
point centers in the coordinate three--dimensional space. In  case of
magnetic, axion and NUT charges absense the represented class transforms
into the earlier obtained by Gibbons \cite {gib}. The connection with other
known special solutions \cite {ren2}, \cite {jm}
is established during the study of the case of null--curvature matrix $M$ linear
dependance on the function $\lambda$. Then
Majumdar--Papapetrou--like black hole solutions family and massless naked
singularities can be obtained from the general one. A comlete list of these
particular solutions is given in the last section of the article.


\section{Matrix Representation of the Stationary String Gravity Equations}

Let us discuss low energy effective four--dimensional action, which
describes the bosonic sector of the heterotic string, taking into account the
contribution of gravitational, Abelian vector, dilaton and axion fields:
\begin{eqnarray}\label{e1}
S = \int d^4x {\mid g \mid}^{\frac {1}{2}} ( && -R+2{\partial \phi}^2+
\frac {1}{2}
e^{4\phi }{\partial \kappa}^2 \nonumber \\ && -e^{-2\phi}F^2 -
\kappa F\tilde {F}),
\end{eqnarray}
where $R=R^{\mu \nu}_{..\mu \nu}$ is the Ricci scalar
$(R^{\mu}_{.\nu \lambda \sigma} =
\partial _{\lambda}\Gamma ^{\mu}_{\nu \sigma}...)$
of the 4-metric $g_{\mu \nu}$, signature $+ - - -$, $\mu = 0,...,3$ and

\begin{eqnarray}\label{e2}
F_{\mu \nu}&=\partial _{\mu}A_{\nu}-\partial _{\nu}A_{\mu},
\nonumber \\
\tilde {F}^{\mu \nu}&=\frac {1}{2} E^{\mu \nu \lambda \sigma}F_{\lambda \sigma}.
\end{eqnarray}

In doing so we consider that the scalar field $\phi$ is the dilaton one, and
the axion is written in the form of pseudoscalar field $\kappa$.

As it has been done \cite{kn}, \cite{iw}, the four--dimensional line
element can be parametrized according to
\begin{equation}\label{e3}
ds^2=f(dt-\omega _idx^i)^2-f^{-1}h_{ij}dx^idx^j,
\end{equation}
where $i=1,2,3$. Below we will study the stationary case when both the metric
and the matter fields are time independent. It has been shown before
\cite {jmp}
that in this case part of the Euler-Lagrange equations can be used for the
transition from both spatial components of the vector potential $A_i$
and entered
in (\ref{e3}) functions $\omega _i$ to the magnetic $u$ and rotation $\chi$
potentials respectively. The new and old variables are connected by
differential relations:
\begin{equation}\label{e4}
\nabla u=fe^{-2\phi}(\sqrt{2}\nabla \times \vec A+\nabla v \times \vec \omega)
+\kappa \nabla v,
\end{equation}
\begin{equation}\label{e5}
\nabla \chi =u\nabla v-v\nabla u -f^2\nabla \times \vec \omega.
\end{equation}
The new notation  $v=\sqrt{2}A_0$ is entered and the three--dimensional operator
$\nabla$ is
corresponded to the three--dimensional metric $h_{ij}$. Also it has been found
that expressed in terms of $f,\chi, u, v, \phi, \kappa$ variational equations
for the action (\ref{e1}) are at the same time Euler-Lagrange equations for the three
dimensional action
\begin{equation}\label{e6}
^3S=\int d^3x h^{\frac {1}{2}}(-^3R+^3L),
\end{equation}
where $^3R$ is the curvature scalar constructed according to 3--metric
$h_{ij}$ and
\begin{eqnarray}\label{e7}
^3L=&\frac {1}{2}f^{-2}[(\nabla f)^2+(\nabla \chi +v\nabla u-u\nabla v)^2]
\nonumber \\
&-f^{-1}[e^{2\phi}(\nabla u-\kappa \nabla v)^2+e^{-2\phi}(\nabla v)^2]
\nonumber \\
&+ 2(\nabla \phi)^2 + \frac {1}{2}e^{4\phi}(\nabla \kappa)^2
\end{eqnarray}
Thus in the stationary case the string gravity appears to be the nonlinear
$\sigma$-model. As it was shown \cite {jmp}--\cite {pr2},
the three--dimensional Lagrangian $^3L$
is invariant  under the ten--parametric continuous transformation
group isomorphic to $Sp(4,R)$. Then it was established that $^3L$ can be
rewritten with the aid of the four--dimensional matrix $M$ in the form
\begin{equation}\label{e8}
^3L=\frac {1}{4}Trj^2,\qquad j=\nabla M M^{-1},
\end{equation}
and $M$, being the matrix of the coset $Sp(4,R)/U(2)$, satisfies the
symplectic and symmetric properties,
\begin{equation}\label{e9}
M^TJM=J,\qquad M^T=M,
\end{equation}
where
\begin{eqnarray}\label{e10}
J=\left (\begin{array}{crc}
0&-I\\
I&0\\
\end{array}\right ).
\end{eqnarray}
It's easy to see that any symplectic matrix $G$ defines automorphism
$M\rightarrow G^TMG$ for the coset under consideration.
The relations (\ref{e9}) allow to parametrize the matrix by six
independent functions
which can be chosen as potentials $f, \chi, u, v, \phi, \kappa$.
Here it is convenient to use the two--dimensional matrices
$P$ and $Q$ which define the Gauss decomposition
\begin{eqnarray}
M=\left (\begin{array}{crc}
P^{-1}&P^{-1}Q\\
QP^{-1}&P+QP^{-1}Q\\
\end{array}\right ).
\end{eqnarray}
Their evident form is \cite {pl}:
\begin{eqnarray}
P=\left (\begin{array}{crc}
f-v^2e^{-2\phi}&-ve^{-2\phi}\\
-ve^{-2\phi}&-e^{-2\phi}\\
\end{array}\right ),
\end{eqnarray}
\begin{eqnarray}
Q=\left (\begin{array}{crc}
-\chi +vw&w\\
w&-\kappa\\
\end{array}\right ),
\end{eqnarray}
where $w=u-\kappa v$.

\section{The General Geodesic Isotropic Solution}

The appropriate to the three--dimensional
action motion equations have the standart form

\begin{equation}
\nabla j=0,
\end{equation}
\begin{equation}
^3R_{ik}=\frac {1}{4}Tr(j_ij_k),
\end{equation}
and admit the procedure of exact solution construction stated before by
Kramer and Neugebauer for arbitrary $\sigma$-models \cite {kn}
and developed later by Clement for the case of $SL(3,R)/SO(2,1)$ matrix
representation of Kaluza--Klein five--dimensional theory
\cite {cl1}--\cite {cl3}.
We consider the ansatz for which the matrix
$M$ is determined by the aid of one space coordinate function $\lambda$
\begin{equation}
M=M(\lambda),\qquad \lambda =\lambda(x^i),
\end{equation}
when $\lambda(x^i)$ is supposed to satisfy the Laplace equation:
\begin{equation}
\nabla ^2\lambda =0.
\end{equation}
It is not difficult to prove that  the `material' equation (14) turns
into a relation, determining the dependance of $M$ on $\lambda$:
\begin{equation}
\frac {d}{d\lambda}\left (\frac {dM}{d\lambda}M^{-1}\right )=0.
\end{equation}
The sense of (18) becomes clear after the introduction of the so called target
space, i.e., of the metric space with the coordinates
$f, \chi, u, v, \phi, \kappa$ and the linear element
uniquely connected with the Lagrangian $^3L$:
\begin{equation}
dS^2=\frac {1}{4}Tr(dM M^{-1} dM M^{-1}).
\end{equation}
Then the formula $M=M(\lambda)$ determines a line in the target space which
according to (18) is a geodesic one \cite {kn}.

The solution of the equation (18) is
\begin{equation}
M=SM_0,
\end{equation}
where
\begin{equation}
S=e^{\lambda A}=\sum _{0}^{\infty}\frac {(\lambda A)^n}{n!},
\end{equation}
$A=const$ and $M_0=M\mid _{\lambda = 0}$.
The matrix $S$, so far is only formally determined,
later on by natural causes it will be called the evolutionary operator.

The three--dimensional Einstein equations (15) in view of (20) and (21)
can be rewritten
\begin{equation}
^3R_{ik}=\frac {1}{4}\lambda _{,i}\lambda _{,k}TrA^2
\end{equation}
and form together with (17) the complete system of equations which determines
the three--dimensional metric $h_{ik}$ and the scalar function $\lambda$.

Let us now establish the conditions for the evolutionary operator $S$
and the matrix $A$ determined by it. Their fulfilment ensures that $M$ belongs
to the coset space $Sp(4,R)/U(2)$ along the whole geodesic line,
only if it is true for  $\lambda =0$, i.e., for the matrix $M_0$.
It is evident that if the matrices $M_0$ and
$M$ are symplectic ones, the operator $S$ should possess the same feature,
and it is convenient to rewrite the first of the relations (\ref{e9}) for it
in the form of
\begin{equation}
S^T=-JS^{-1}J,
\end{equation}
hence $A$ can be immediately determined:
\begin{equation}
A^T=JAJ.
\end{equation}
Thus, $A$ is an element of $sp(4,R)$ algebra; and after the
solution of (24), it can be represented as
\begin{eqnarray}
A=\left (\begin{array}{crc}
-s^T&r\\
l&s\\
\end{array}\right ),
\end{eqnarray}
where $l^T=l,\quad r^T=r$ and $s$ are the two--dimensional matrices
which in sum define ten independent parameters.

Then, in order that $M^T=M$ followed from $M_0^T=M_0$, the evolutionary
operator should satisfy the (nongroup) condition
\begin{equation}
S^T=M_0^{-1}SM_0,
\end{equation}
which imposes on $A$ the following restriction:
\begin{equation}
A^T=M_0^{-1}AM_0.
\end{equation}
Here we can finish the general analysis and turn our attention to the
solutions determined by isotropic geodesic lines in the target space.
From (19)--(21) follows that condition $dS^2=0$ is equivalent to
\begin{equation}
TrA^2=0.
\end{equation}
Then from (17) and  (22) immediately follows that  $h_{ik}$ and $\lambda (x^i)$
can be taken in the form
\begin{equation}
h_{ik}=\delta _{ik},\qquad
\lambda =\sum \frac {\lambda _n}{\mid \vec r-\vec r_n\mid}
\end{equation}
where $\vec r$ is as usual connected to $x^i$ and $\vec r_n$ is considered
as the position of the n--center characterized by $\lambda _n$.
We will assume that $\sum \lambda _n \ne 0$ (the dropped
special case can be investigated in the same way), then in view of (21),
without loss of generality, it is possible to impose on
$\lambda _n$ the normalization condition
\begin{equation}
\sum _{n} \lambda _n=1.
\end{equation}
It is evident that $\lambda \rightarrow 0$ when $r\rightarrow \infty$,
thus $M_0=M_{\infty}$. Let us naturally determine
the asymptotic values of physical fields assuming that
\begin{equation}
f_{\infty}=1,\quad  \chi _{\infty}=u_{\infty}=v_{\infty}=\phi _{\infty}=
\kappa _{\infty}=0;
\end{equation}
then from (11)--(13) we obtain
\begin{eqnarray}
M_0=\left (\begin{array}{crc}
\sigma _3&0\\
0&\sigma _3\\
\end{array}\right )\equiv \Sigma _3,
\end{eqnarray}
and $\sigma _3$ is one of the Pauli matrices. By substituting the found
value $M_0$ in (27), $A$ can be calculated as
\begin{eqnarray}
A=\left (\begin{array}{crc}
-\tilde s&r\\
\tilde r&s\\
\end{array}\right ),
\end{eqnarray}
where $s^T=\tilde s$ and for any two--dimensional matrix $m$ we define
$\tilde m=\sigma _3m\sigma _3$.

From (21), (29), (30) it follows that at $r\rightarrow \infty$
\begin{equation}
S\longrightarrow I+\frac {A}{r}
\end{equation}
and because of (20)
\begin{equation}
M\longrightarrow \Sigma _3+\frac {A\Sigma _3}{r}.
\end{equation}
After that, applying (11)--(13) and (25) it is easy to show that
the main parts of the asymptotic decomposition of the functions
$f-1, \chi, u, v, \phi, \kappa$ are proportional to $\frac {1}{r}$. In this
case, six components of matrices $s$ and $r$ act as coefficients which
in this way determine six physical charges of the system. By entering the mass
$M$, the parameter NUT $N$ and also the electric $Q_e$, magnetic $Q_m$,
dilaton $D$ and axion $A$ charges according to formulae
\begin{eqnarray}
f& \rightarrow  1-\frac{2M}{r},\qquad \chi \rightarrow \frac{2N}{r},
\nonumber \\
v&  \rightarrow  \frac{\sqrt 2 Q_e}{r},\qquad u \rightarrow
\frac{\sqrt 2 Q_m}{r}&,
\nonumber \\
\phi & \rightarrow \frac{D}{r},\qquad \kappa \rightarrow \frac{2A}{r},
\end{eqnarray}
$s$ and $r$ are found as follows:
\begin{eqnarray}
s=\left (\begin{array}{crc}
-2M&\sqrt 2Q_e\\
-\sqrt 2Q_e&-2D\\
\end{array}\right ),
\noindent \\
r=\left (\begin{array}{crc}
2N&-\sqrt 2Q_m\\
-\sqrt 2Q_m&-2A\\
\end{array}\right ).
\end{eqnarray}

Let us determine now the evident form of the evolutionary operator $S$, which
was written before with the aid of the formal exponential
series. The calculation
of $A^2$ in view of (33) gives:
\begin{equation}
A^2=\alpha ^{\mu}T_{\mu},
\end{equation}
where parameters $\alpha ^{\mu}$ are of the second order with respect
to charges, $T_0$ is
the unit matrix and three traceless matrices $T_i$ are
\begin{eqnarray}
T_1=\left (\begin{array}{crc}
\sigma _2&0\\
0&-\sigma _2\\
\end{array}\right ),
\qquad
T_2=\left (\begin{array}{crc}
0&\sigma _2\\
\sigma _2&0\\
\end{array}\right ),
\qquad
T_3=\Sigma _3.
\end{eqnarray}

It is convenient to unite six real charges into three complex ones:
\begin{eqnarray}
{\cal M}&=M+iN,
\nonumber \\
{\cal D}&=D+iA,
\nonumber \\
{\cal Q}&=Q_e+iQ_{m},
\end{eqnarray}
in terms of which
\begin{eqnarray}
\alpha ^0&=&2(\bar {\cal M}{\cal M}+\bar {\cal D}{\cal D}-
\bar {\cal Q}{\cal Q}),
\nonumber \\
\alpha ^1+i\alpha ^2&=&-2\sqrt 2({\cal M}\bar {\cal Q}+\bar {\cal D}{\cal Q}),
\nonumber \\
\alpha ^3&=&2(\bar {\cal M}{\cal M}-\bar {\cal D}{\cal D}).
\end{eqnarray}

In doing so the isotropic condition (28) can be
rewritten as
\begin{equation}
\bar {\cal M}{\cal M}+\bar {\cal D}{\cal D}=\bar {\cal Q}{\cal Q}
\end{equation}
and generalizes the  known relations in the Einstein-Maxwell theory
\cite {per}--\cite {m}.

It is easy to verify that the commutators of the matrices $T_i$ are
not their linear
combinations, i.e., these matrices do not form the basis of a three--dimensional
Lie algebra. But the calculation of the corresponding anticommutators leads
to the following result:
\begin{equation}
\{ T_i,T_j\} =T_iT_j+T_jT_i=-\eta _{ij},
\end{equation}
where $\eta _{ij}=diag(1, 1, -1)$, thus in view of (39) and (40) it appears that
\begin{equation}
A^4=-\eta _{ij}\alpha ^i\alpha ^j.
\end{equation}
The application of the relation (43) also allows to determine the fact that the
quadratic form $\eta _{ij}\alpha ^i\alpha ^j$ is not negative and enter
a new parameter $\alpha$ according to the definition
\begin{equation}
\alpha ^4=\eta _{ij}\alpha ^i\alpha ^j
\end{equation}
The relation (46) allows to sum the series which correspond to the items with
$n=4k$ from the exponent decomposition (21). Now it is not difficult to find the
evident form for the remaining three series with
$n=4k+1$, $n=4k+2$ and $n=4k+3$. The mentioned four
series just compose the evolutionary operator and its expression in
terms of the charge matrix $A$, defined by
(33), (37), (38) and (43), and by the
function $\lambda$ (29) is
\begin{equation}
S=\sum _0^3S_{\mu}A^{\mu},
\end{equation}
where $A^{\mu}$ is the matrix $A$ to the $\mu$ power and
\begin{eqnarray}
S_0&=&\cosh (\alpha \lambda / \sqrt 2)\cos (\alpha \lambda / \sqrt 2),
\nonumber \\
S_1&=\frac {1}{\sqrt 2\alpha}
&(\sinh (\alpha \lambda / \sqrt 2)\cos (\alpha \lambda / \sqrt 2)
\nonumber \\
&+&\cosh (\alpha \lambda / \sqrt 2)\sin (\alpha \lambda / \sqrt 2)),
\nonumber \\
S_2&=\frac {1}{\alpha ^2}
&\sinh (\alpha \lambda / \sqrt 2)\sin (\alpha \lambda / \sqrt 2),
\nonumber \\
S_3&=\frac {1}{\sqrt 2\alpha ^3}
&(\cosh (\alpha \lambda / \sqrt 2)\sin (\alpha \lambda / \sqrt 2)
\nonumber \\
&-&\sinh (\alpha \lambda / \sqrt 2)\cos (\alpha \lambda / \sqrt 2)).
\end{eqnarray}

The constructed solution (29) and (47)--(48)
defines the system of interacting point centers which satisfies
the restriction (43).

Let us turn our attention to the group nature of the matrix $S$. From (33)
it is evident that the determining $S$ matrix $A$ differs from the belonging
to the $sp(4,R)$ algebra general matrix by
\begin{eqnarray}
\Gamma =\left (\begin{array}{crc}
\tilde \tau &\rho \\
-\tilde \rho &\tau \\
\end{array}\right ),
\end{eqnarray}
where $\tilde \tau =-\tau ^T$, and $\rho$ is symmetric.
Entered here $\Gamma$ has four independent parameters, the corresponding linear
independent matrices (generators) can be written as
\begin{eqnarray}
\Gamma _0=\left (\begin{array}{crc}
0&\sigma _3\\
-\sigma _3&0\\
\end{array}\right ),
\Gamma _1=\left (\begin{array}{crc}
0&\sigma _1\\
\sigma _1&0\\
\end{array}\right ),
\nonumber \\
\Gamma _2=\left (\begin{array}{crc}
-\sigma _1&0\\
0&\sigma _1\\
\end{array}\right ),
\Gamma _3=\left (\begin{array}{crc}
0&I\\
-I&0\\
\end{array}\right ).
\end{eqnarray}
It is easy to prove that
\begin{equation}
[\Gamma _0, \Gamma _i]=0
\end{equation}
and pair products for $\Gamma _i$ are
\begin{equation}
\Gamma _i\Gamma _j=I\eta _{ij}+\epsilon _{ijk}\eta ^{kl}\Gamma _l
\end{equation}
where $\eta ^{kl}=\eta _{kl}$.

Because of the resulting from (52) commutation relations,
the isomorphism between
algebra of $\Gamma _i$ and that of two--dimensional Pauli matrices $\sigma _i$
\begin{eqnarray}
\sigma _1=\left (\begin{array}{crc}
0&1\\
1&0\\
\end{array}\right ),
\sigma _2=\left (\begin{array}{crc}
0&-1\\
1&0\\
\end{array}\right ),
\sigma _3=\left (\begin{array}{crc}
1&0\\
0&-1\\
\end{array}\right ),
\end{eqnarray}
belonging to $sl(2,R)$, can be determined as
\begin{equation}
\Gamma _1\sim \sigma _1,\quad \Gamma _2\sim \sigma _3,\quad
\Gamma _3\sim \sigma _2.
\end{equation}
If we also identify $\Gamma _0\sim I$,
it is easy to notice that the algebra of matrices $T_{\mu}$
appears to be isomorphic to $sl(2,R)\bigoplus R \sim gl(2,R)$.
So the part of the group omitted while constructing the evolutionary operator
is locally isomorphic to $GL(2,R)$ and hence
$S\in Sp(4,R)/GL(2,R)$.

Later on it can be seen that the matrix
\begin{equation}
G=e^{\Gamma}=e^{\gamma ^{\mu}\Gamma _{\mu}}
\end{equation}
in view of (49) satisfies the relation
\begin{equation}
G^T\Sigma _3G=\Sigma _3.
\end{equation}
This allows to interpret $G$ as a general matrix of belonging to $Sp(4,R)$
transformation which preserves the asymptotical vacuum values for the
system of physical fields.

It is necessary to remark that the formal expression (55) as the corresponding
one for $S$ can be obtained in the evident form. Namely, let us denote the
matrix constructed on $\Gamma _0$ by $G_{(0)}$ and that of constructed on
$\Gamma _i$ by $G_{(3)}$.
Then in view of (51)
\begin{equation}
G=G_{(0)}G_{(3)}=G_{(3)}G_{(0)}.
\end{equation}
Employing the relation
\begin{equation}
\Gamma _0^2=-I
\end{equation}
we have that
\begin{equation}
G_{(0)}=I\cos \gamma ^0+\Gamma _0\sin \gamma ^0.
\end{equation}
Then, noticing that from (52) follows the anticommutation relations
$\{ \Gamma _i\Gamma _j\} =2I\eta _{ij}$
the expression for $G_{(3)}$ matrix can be found:
\begin{eqnarray}
2G_{(3)}=I&[(1+\sigma)\cosh \gamma +(1-\sigma)\cos \gamma]
\nonumber \\
+\frac {\gamma ^i\Gamma _i}{\gamma}
&[(1+\sigma)\sinh \gamma +(1-\sigma)\sin \gamma]
\end{eqnarray}
where the parameter $\gamma$ is determined as
\begin{equation}
(\gamma)^2=\sigma \eta _{ij}\gamma ^i \gamma ^j
\end{equation}
and $\sigma =sign(\eta _{ij}\gamma ^i \gamma ^j)$.


\section{Black Holes and Naked Singularities}

The general geodesic isotropic solution of the string gravity obtained
in the previous part has intricate dependence from the function $\lambda$,
which satisfies Laplace equation and hence, from the space coordinates.
It is easy to verify that in the case when $\alpha=0$, i.e., if
\begin{equation}
\alpha_i\alpha_j\eta^{ij}=0
\end{equation}
the evolutionary operator $S$ becomes the polynomial of third power on
$\lambda$ which considerably facilitates the solution analysis. The greatest
simplification is obtained at the simultaneous imposure of the
set of three additional relations $\alpha^i=0$
to the physical
charges, which according
to (42) are equivalent to
\begin{eqnarray}
{\cal M}\bar {\cal Q}+{\cal Q}\bar {\cal D}=0,
\nonumber \\
{\cal M}\bar {\cal M}-{\cal D}\bar {\cal D}=0.
\end{eqnarray}
In doing so, in view of (21) and (39) the evolutionary operator
and the null-curvature
matrix $M$ occure to be the linear functions of $\lambda$ and satisfy,
according to (17), the Laplace equation. The result can be investigated and,
as it is further demonstrated, it contains interesting physical solutions.
At first and foremost the independent `coordinates' can be entered in the
charge space. The number of such `coordinates' appears to be equal to three,
as it immediately follows from (43) and (63). By applying the complex form of
transcription (41) we have:
\begin{eqnarray}
{\cal M} &=& \rho e^{2i\delta_1},\qquad
{\cal D} = \rho e^{2i\delta_2},
\nonumber \\
{\cal Q} &=& -i\sqrt{2}\sigma\rho e^{i(\delta_1+\delta_2)},
\end{eqnarray}
or, going to the real charges
\begin{eqnarray}
M & = & \rho \cos{2\delta_1}, \nonumber \\
N & = & \rho \sin{2\delta_1}, \nonumber \\
D & = & \rho \cos{2\delta_2}, \nonumber \\
A & = & \rho \sin{2\delta_2},  \\
Q_e & = & \sqrt{2}\sigma \rho \sin{(\delta_1-\delta_2)}, \nonumber \\
Q_m & = & -\sqrt{2}\sigma \rho \cos{(\delta_1-\delta_2)}, \nonumber
\end{eqnarray}
and $\sigma =\pm 1$.
Turning back to relation (20) it is possible to calculate the matrix $M$.
After that, employing the Gauss decomposition formula (11) and formulae (12)
and (13),
which determine the explicit dependance of matrix elements from six
independent $\sigma$--model functions, the expressions can be found as follows:
\begin{eqnarray}
f & = & (1+2M\lambda)^{-1}, \nonumber \\
\chi & = & -2N\lambda (1+2M\lambda)^{-1}, \nonumber \\
v & = & \sqrt{2}Q_e\lambda (1+2M\lambda)^{-1}, \nonumber \\
u & = & \sqrt{2}Q_m\lambda (1+2M\lambda)^{-1}.
\end{eqnarray}
In this case the expressions for axion and dilaton appear to be rather
cumbersome, but by entering the complex variable
\begin{equation}
z=\kappa+ie^{-2\phi}
\end{equation}
which, according to \cite {pl}, is one of the Ernst potentials
for the stationary
system (\ref{e6})--(\ref{e7}) the following compact result, which generalizes expressions
found in \cite {ren1} and \cite {ren2}, can be obtained
\begin{equation}
z=i\frac{1+\lambda({\cal M}-{\cal D})}{1+\lambda({\cal M}+{\cal D})}.
\end{equation}
In this case, when the solution is determined only by one center, i.e., when
$\lambda=\frac{1}{r}$, it is convenient to turn to a new radial coordinate
$R=r+2M$. It is easy to show that the expressions for electric and magnetic
potentials transform to the most simple Coulomb form
\begin{eqnarray}
v &=& \frac{\sqrt{2}Q_e}{R},\qquad u=\frac{\sqrt{2}Q_m}{R},\nonumber \\
z &=& i\frac{R-\bar {\cal M}-{\cal D}}{R-\bar {\cal M}+{\cal D}}.
\end{eqnarray}
The employment of (\ref{e5}) and (66) allows to determine the obvious form of
four--dimensional space--time metric:
\begin{eqnarray}
ds^2 & = & (1-\frac{2M}{R})(dt-2Ncos{\theta } d\phi)^2
\nonumber \\
& - & (1-\frac{2M}{R})^{-1}dR^2 - R(R-2M)d\Omega ^2.
\end{eqnarray}
In the constructed solution the mass, which causes the horizon appearance at
$R=R_H=2M$, and the parameter NUT which makes the spatial interval
asymptotically
different from Minkowski metric, appear to be independent parameters.
This means that exist special solutions, which are asimptotically flat and
have the
horizon (black holes) and also solutions, possessing everywhere the regular
but asimptotically unflat four--dimensional metric with the Coulomb-like
expressions for the material fields (the naked singularities). It is important
to note that the presence of the naked singularities in the string gravity
appears to be possible due to the existence of the scalar sector in the
theory, i.e., dilaton and axion fields.

From here on while investigating the special cases, both the results for the
multicenter system and formulae (69) and (70) describing isolated
singular object,
will be taken into account.
So, let us discuss the case $N=0$, that according to (65) and condition $M>0$ is
equivalent to the relation $\delta_1=\pi n$. The corresponding formulae for
the charges lead to the following expressions:
\begin{equation}
D \sim Q_m^2-Q_e^2, \qquad A \sim Q_mQ_e
\end{equation}
It is seen that the dilaton black holes (with $A=0$) have one of the
electromagnetic charges equal to zero while the axion black holes
(for them $D=0$) have equal in absolute magnitude electric and magnetic
charges. Hence from the constructed before family of solutions (66), (68) and
(69)--(70)
naturally four black holes subfamilies stand out:
dilaton magnetic, for which $\delta_2=\pi k$,
\begin{equation}
D=M,\quad  Q_m=\sqrt{2}\sigma M,
\end{equation}
and all the other charges are equal to zero;
dilaton electric, for which $\delta_2=\pi (k+1/2)$
\begin{equation}
D=-M,\quad  Q_e=\sqrt{2}\sigma M;
\end{equation}
axion with $Q_e=-Q_m$ when $\delta_2 =\pi (k+1/4)$
\begin{equation}
A=M,\quad  Q_e=-Q_m=\sigma M;
\end{equation}
and, at last, axion with $Q_e=Q_m$ appearing at
$\delta_2=\pi (k+3/4)$ and possessing
\begin{equation}
A=-M,\quad  Q_e=Q_m=\sigma M.
\end{equation}
One can notice that the discret transformation $Q_m\rightarrow Q_e,\quad
Q_e\rightarrow -Q_m$ transforms magnetic dilaton solution (72) into
electric one (73) with simultaneous change $D\rightarrow -D$. In doing
so the appropriate axion solutions transform one into another, if
besides of the above mentioned electromagnetic transformation
$A\rightarrow -A$ taking place.

Now we can study the massless solution families having, in accordance with (65)
the NUT parameter value not equal to zero. From (70) it can be seen that
the space metric is regular everywhere and only matter
fields have physical peculiarities.
So let us examine the case $M=0$. As the parameter $N$ can be of different
sign, from (65) follows that $\delta_1 =\pi (2n+1)/4$. Omitting the
technical details we will turn our attention to the discussion of the main
results. It turns out that the family of naked singularities under
investigation, just as described above the black holes family, has the four
most simple solution classes. Namely, the case when $\delta_1 =\pi (l+1/4),
\delta_2=\pi (k+1/4)$ and $\delta_1 =\pi (l+3/4), \delta_2 =\pi (k+3/4)$
corresponds to the axion magnetic solution (constructed earlier in
\cite {pr2}):
\begin{equation}
A=N,\quad Q_m=\sqrt{2}\sigma N;
\end{equation}
when $\delta_1=\pi (l+1/4), \delta_2=\pi (k+3/4)$ and
$\delta_1=\pi (l+3/4), \delta_2=\pi (k+1/4)$ we get
dilaton electric singularity:
\begin{equation}
A=-N,\quad Q_e=\sqrt{2}\sigma N;
\end{equation}
when $\delta_1=\pi (l+3/4), \delta_2=\pi (k+1/2)$ and $\delta_1=\pi (l+1/4),
\delta_2= \pi k$ it appears that
\begin{equation}
D=N,\quad Q_e=-Q_m=\sigma N,
\end{equation}
and we get dilaton singularity with electromagnetic charges of different sign;
and, finally, when $\delta_1 = \pi (l+1/4), \delta_2 =\pi (k+1/2)$ and
$\delta_1=\pi (l+3/4), \delta_2 =\pi k$
the fields configuration of
the dilaton singularity is determined by equal values of electric and magnetic
charges, which are connected with NUT parameter as follows:
\begin{equation}
D=-N,\quad Q_e=Q_m=\sigma N.
\end{equation}

Similarly to the situation with black holes, the determined above discret
transformations acting in the charge space connect axion singularities (76)
and (77) and also (with the corresponding change of axion charge to dilaton one)
transfer dilaton solutions (78)--(79) one into another. It is important to
stress the resulting from formulae (72)--(75)
and (76)--(79) formal analogy between
dilaton black holes and naked axion singularities from one hand and axion
black holes and dilaton naked singularities from the other.
Turning back to the solution (65), (69), (70) describing singular object
with mass
and parameter NUT it can be pointed out that the solutions describing
asimptotic flat black holes can be transformed into solutions for the
horizonless asymptotic nonflat naked singularities with the aid of continuous
transformation of the parameter $\delta_1$.


\section{Conclusion}
Using the Kramer--Neugebauer method for the null--curvature matrix
Sp(4,R)/U(2) coset formulation of the stationary D=4 EMDA theory we
have constructed a new class of solutions which describe a system of interacting
point centers. These centers describe a set of Majumdar--Papapetrou--like
black holes in a special case and massless naked singularities in another one.
A general class is connected with a complete family of isotropic geodesic
lines which are crossing in a Minkowski vacuum point of the target space.
As it has
been shown, the evolutionary operator transforming vacuum solution to nontrivial
one belongs to Sp(4,R)/GL(2,R) coset. The ommited four generators of GL(2,R)
subgroup defines the general automorphism for the Sp(4,R)/U(2) target space
which preserves asymptotic flatness.

Used formalism admits the natural generalization when solutions are defined by
extremal area surfaces in the potential space. It gives the possibility to
construct Israel--Wilson--like sourses for the theory under consideration.


\acknowledgments

This work was supported in part by the ISF Grant No. M79000.


\end{document}